\documentclass[12pt,a4,fullpage,amssymbols]{article}
\usepackage{amssymb, amsmath}
\renewcommand{\theequation}{\arabic{section}.\arabic{equation}}

\begin{document}
\title{An exact and explicit formula for pricing Asian options with regime switching}
\author{ Leunglung Chan\footnote{School of Mathematics and Statistics, University of New South Wales,
Sydney, NSW, 2052, Australia, Email: leung.chan@unsw.edu.au}\enspace and
Song-Ping Zhu\footnote{School of Mathematics and Applied Statistics, University of Wollongong, Wollongong,
NSW 2522, Australia, Email: spz@uow.edu.au}}
\maketitle
\begin{abstract}
This paper studies the pricing of European-style Asian
options when the price dynamics of the underlying risky asset are
assumed to follow a Markov-modulated geometric Brownian motion; that is, the appreciation rate and the volatility of the
underlying risky asset depend on unobservable states of the economy
described by a continuous-time hidden Markov process.
We derive the exact, explicit and closed-form solutions for European-style Asian options in a two-state regime switching model.
\end{abstract}
{\bf Key words:}
Option pricing; Markov-modulated geometric Brownian motion; Regime switching;
Asian options.
\section{Introduction}
The pricing, hedging and risk management of contingent claims has become a popular topic because contingent claims are now widely used to transfer risk in financial markets.
The pioneering work of Black and Scholes (1973) and Merton (1973) laid the foundations
of the field and stimulated important research in option pricing theory.
The Black-Scholes-Merton formula has been widely adopted by traders,
analysts and investors. The contingent claims traded in the market not only include vanilla European options but also exotic options such as Asian options.
Asian options are path-dependent options whose payoff depends on an average of the stock prices over a certain time period. Asian options are used for hedging purpose. Traders may be interested to hedge against the average price of a commodity over a period. The use of Asian options may avoid price manipulation near the end of the period. Most Asian options are European style because an Asian option with the American early exercise feature may be redeemed as early as the beginning of the averaging period and thus lose hedging purpose from averaging. Since no general closed-form solution for the price of the Asian option based on arithmetic averaging is known, a variety of methods have been developed to study this problem. Many papers studied Asian options including Kemna and Vorst (1990), Turnbull and
Wakeman (1991), Ritchken et al. (1993), Geman and Yor (1993), Curran (1994), Rogers and Shi (1995), Zhang
(1995),  Boyle et al. (1997) and Fu et al. (1999).

Despite its popularity the Black-Scholes-Merton model fails in various ways, such as the fact that implied volatility is not constant. During the past few decades many extensions to the Black-Scholes-Merton model have been
introduced in the literature to provide more realistic descriptions for asset price dynamics. In particular, many models have been introduced
to explain the empirical behavior of the implied volatility smile and smirk. Such models include the stochastic volatility models,
jump-diffusion models, models driven by L\'{e}vy processes and regime switching models.

Maybe the simplest way to introduce additional randomness into the standard
Black-Scholes-Merton model is to let the volatility and rate of return be functions of a finite state
Markov chain. There has been considerable interest in applications of regime
switching models driven by a Markov chain to various financial
problems. Many papers in a regime switching framework include Elliott and van der Hoek (1997),
 Guo (2001), Elliott et al. (2001), Buffington and Elliott (2002a,b), Elliott et al. (2003) and Elliott et al. (2005). In addition, Boyle and Draviam (2007) used the
numerical method to solve the system of coupled partial differential equations for the price of exotic options under regime switching.
Siu et al. (2008) priced credit default swaps under a Markov-modulated
Merton structural model. Yuen and Yang (2009) proposed a recombined trinomial tree to price simple options and barrier options in a jump-diffusion model with regime switching. Yuen and Yang (2010) used a trinomial tree method to price Asian options and equity-indexed annuities with regime switching. Zhu et al. (2012) derived a closed-form solution for European options with a two-state regime switching model. However, there is no closed-form solution to Asian options under a regime switching model.

In this paper, we study the pricing of European-style
Asian options when the price dynamics of the underlying
risky asset are assumed to follow a Markov-modulated geometric Brownian motion.
The Markov-modulated
model provides a more realistic way to describe and explain the market
environment. It has been mentioned in Yao et al. (2003) that
it is of practical importance to allow the market
parameters to respond to the movements of the general market levels
since the trend of general market levels is a key factor which
governs the price movements of individual risky assets.
We derive an analytical solution for Asian option
by means of the homotopy analysis method (HAM).
HAM was initially suggested by Ortega and Rheinboldt (1970) and has been successfully used to solve a number of heat transfer problems, see Liao (1997), Liao and Zhu (1999), and fluid-flow problems, see Liao and Zhu (1996), Liao and Campo (2002). Zhu (2006) used HAM to obtain an analytic pricing formula for American options in the Black-Scholes model. Gounden and O'Hara (2010) extended the work of Zhu to pricing  American-style Asian options of floating strike type in the Black-Scholes model. Leung (2013) used HAM to derive an analytic formula for lookback options under stochastic volatility.

This paper is organized as follows.
Section $2$ describes the asset price dynamics under the Markov-modulated geometric Brownian motion.
Section $3$ formulates the partial differential equation system for the price of a floating-strike Asian option. Section $4$ derives an exact, closed-form solution for the floating-strike Asian option. Section $5$ discusses a symmetry between fixed-strike and floating-strike Asian options. The final section contains a conclusion.
\section{Asset Price Dynamics}
Consider a complete probability space $(\Omega, {\cal F}, {\cal P})$,
where $\cal P$ is a real-world probability measure.
Let $\cal T$ denote the time index set $[0,T]$  of the model.
Write $\{W_t\}_{t \in \cal T}$ for a standard Brownian motion on
$(\Omega, {\cal F}, {\cal P})$. Suppose the states of an economy
are modelled by a finite state continuous-time Markov chain
$\{X_t\}_{t \in \cal T}$ on $(\Omega, {\cal F}, {\cal P})$. Without loss of generality, we
can identify the state space of $\{X_t\}_{t \in \cal T}$
with a finite set of unit vectors ${\cal X} := \{e_1, e_2, \dots, e_N\}$,
where $e_i = (0, \dots, 1, \dots, 0) \in {\mathbb{R}}^{N}$.
We suppose that $\{X_t\}_{t \in \cal T}$ and $\{W_t\}_{t \in \cal T}$
are independent.

Let $\tilde{A}$ be the generator $[a_{ij}]_{i,j =1, 2, \dots, N}$ of the
Markov chain. From Elliott et al. (1994), we
have the following semimartingale representation theorem for
$\{X_t\}_{t \in \cal T}$:
\renewcommand{\theequation}{2.\arabic{equation}}
\setcounter{equation}{0}
\begin{eqnarray}
X_t = X_0 + \int^{t}_{0} {\tilde{A}} X_s d s + M_t \ ,
\end{eqnarray}
where $\{M_t\}_{t \in \cal T}$ is an ${\mathbb{R}}^N$-valued
martingale increment process with respect to the filtration
generated by $\{X_t\}_{t \in \cal T}$.

We consider a financial model with two primary traded assets, namely
a money market account $B$ and a risky asset or stock $S$. Suppose
the market is frictionless; the borrowing and lending interest rates
are the same; the investors are price-takers.

The instantaneous market interest rate $\{r(t,
X_t)\}_{t \in \cal T}$ of the bank account is given by:
\begin{eqnarray}
r_t := r(t, X_t) = <r, X_t> \ ,
\end{eqnarray}
where $r := (r_1, r_2, \dots, r_N)$ with $r_i > 0$ for each $i= 1, 2, \dots, N$ and $<\cdot, \cdot>$ denotes the inner product in $\mathbb{R}^{N}$.

In this case, the dynamics of the price process $\{B_t\}_{t \in
\cal T}$ for the bank account are described by:
\begin{eqnarray}
d B_t = r_t B_t dt \ ,  \quad   B_0 = 1 \ .
\end{eqnarray}
Suppose the stock appreciation rate $\{\mu_t\}_{t \in \cal
T}$ and the volatility $\{\sigma_t\}_{t \in \cal T}$ of $S$
 depend on $\{X_t\}_{t \in \cal T}$ and are described by:
\begin{eqnarray}
\mu_t := \mu (t, X_t) = <\mu, X_t> \ , \quad \ \sigma_t := \sigma (t, X_t) = <\sigma,
X_t> \ ,
\end{eqnarray}
where $\mu := (\mu_1, \mu_2, \dots, \mu_N)$, $\sigma := (\sigma_1,
\sigma_2, \dots, \sigma_N)$ with $\sigma_i > 0$ for each $i = 1, 2,
\dots, N$ and $<\cdot, \cdot>$ denotes the inner product in ${\cal
R}^{N}$.

We assume that the price dynamics of the underlying risky asset
$S$ are governed by the Markov-modulated geometric Brownian motion
:
\begin{eqnarray}
d S_t = \mu_t S_t d t + \sigma_t S_t d W_t \ ,
\quad
S_0 = s_{0}.
\end{eqnarray}
\section{Asian option}
\renewcommand{\theequation}{3.\arabic{equation}}
\setcounter{equation}{0}
We now turn to the pricing of an Asian option of floating strike type in a regime switching model. We consider continuously sampled arithmetic average. The average from continuous sampling is given by
\begin{eqnarray*}
\frac{1}{T}\int_{0}^{T}S_{u}du.
\end{eqnarray*}
The price of Asian options based on an arithmetic averaging is not known in closed form even in the Black-Scholes-Merton model.

 We assume that $Q$ is a risk-neutral measure and the price dynamics
of the risky stock under $Q$ are governed by
\begin{eqnarray}
d S_t = r_t S_t d t + \sigma_t S_t d{\tilde{W}}_t \ ,
\quad
S_0 = s_{0}.
\end{eqnarray}
Now define a process
\begin{eqnarray}
A_{t}=\int_{0}^{t}S_{u}du.
\end{eqnarray}
We consider a floating-strike Asian put option whose payoff at time $T$ is
\begin{eqnarray}
 V(T)=\bigg(\frac{1}{T}A_{T}-S_{T}\bigg)^{+}.
\end{eqnarray}
The price at times $t$ prior to the expiration time $T$ of this put is given by the risk-neutral pricing formula
\begin{eqnarray}
V(t,s,a,{\bf x})&=&E^{Q} [ e^{-\int_{t}^{T} r_{u}du}V(T)  | S_{t} = s, A_{t}=a, {\bf X}_{t} = {\bf x} ]\nonumber\\
&=& E^{Q}\bigg [ e^{-\int_{t}^{T} r_{u}du}\bigg(\frac{1}{T}A_{T}-S_{T}\bigg)^{+}| S_{t} = s, A_{t}=a, {\bf X}_{t} = {\bf x} \bigg]\ .
\end{eqnarray}
Write ${\bf V}=(V(t,s,a,e_{1}),...,V(t,s,a,e_{N})).$
Applying the Feynman-Kac formula to the above equation, then $V (t, s,a,{\bf x})$ satisfies the system of partial differential equations
\begin{eqnarray}
\frac{\partial V}{\partial t}+r_{t}s\frac{\partial V}{\partial s}+s\frac{\partial V}{\partial a}+\frac{1}{2}\sigma_{t}^{2}s^{2} \frac{\partial^{2}V}{\partial s^{2}}-r_{t}V+ \left <{\bf V}, {\bf \tilde{A}} {\bf x} \right > = 0 \ ,
\end{eqnarray}
and the boundary conditions
\begin{eqnarray}
V(T,s,a,{\bf x})=\big(\frac{a}{T}-S_{T}\big)^{+},
\end{eqnarray}
\begin{eqnarray}
V(t,s,0,{\bf x})=0.
\end{eqnarray}
The price of an Asian option can be found by solving above system of partial differential equations (PDE)
in two space dimensions. In the case of the Black-Scholes-Merton model, Ingersoll (1987) observed that the three-dimensional PDE for a floating strike Asian option can be
reduced to a two-dimensional PDE. However, by using $\frac{S_{t}}{A_{t}}$ as the state variable, and hence a nontraded variable as the num\'{e}raire, martingale pricing techniques cannot be exploited.
On the other hand, we will adopt $S_{t}$ as the num\'{e}raire and by the change of measure reduce the three-dimensional problem (3.4) to a two-dimensional problem (for instance see Peskir and Shiryaev (2006)).

Write $\{{\cal F}^X_t\}_{t \in \cal T}$ and $\{{\cal F}^W_t\}_{t
\in \cal T}$ for the $\cal P$-augmentation of the natural
filtrations generated by $\{X_t\}_{t \in \cal T}$ and
$\{W_t\}_{t \in \cal T}$, respectively. For each $t \in \cal T$,
we define ${\cal G}_t$ as the $\sigma$-algebra ${\cal F}^X_t \vee
{\cal F}^W_t$.
We introduce a second process
\begin{eqnarray}
Y_{t}=\frac{A_{t}}{S_{t}}
\end{eqnarray}
and define
\begin{eqnarray}
\frac{dQ^{*}}{dQ}|_{{\cal G}_{T}}=e^{-\int_{0}^{T}r_udu}\frac{S_{T}}{S_{0}}
\end{eqnarray}
so that $W^{Q^*}_{t}={\tilde{W}}_{t}-\int_{0}^{t}\sigma_{u}du$ is a standard Brownian motion with respect to $\{{\cal G}_t\}_{t
\in \cal T}$ under $Q^{*}$
and using $S_{t}$ as the num\'{e}raire, the valuation problem (3.4) becomes
\begin{eqnarray}
V(t,y,{\bf x})=E^{Q^{*}}\bigg [\bigg(\frac{1}{T}Y_{T}-1\bigg)^{+}| Y_{t}=y, {\bf X}_{t} = {\bf x} \bigg] \ .
\end{eqnarray}
Here
\begin{eqnarray}
dY_{t}=\big(1-r_{t}Y_{t}\big)dt+\sigma_{t} Y_{t}d{\hat{W}}_{t}^{Q^*},
\end{eqnarray}
where ${\hat{W}}_{t}^{Q^*}=-W^{Q^*}_{t}$ is a standard Brownian motion with respect to $\{{\cal G}_t\}_{t
\in \cal T}$ under $Q^{*}$.
Then via the Feynman-Kac formula, the price of the Asian option $V(t,y,{\bf x})$ satisfies the following system of PDEs:
\begin{eqnarray}
\frac{\partial V}{\partial t}+(1-r_{t}y)\frac{\partial V}{\partial y}+\frac{1}{2}\sigma_{t}^{2}y^{2} \frac{\partial^{2}V}{\partial y^{2}}+ \left <{\bf V}, {\bf \tilde{A}} {\bf x} \right > = 0 \ ,
\end{eqnarray}
and the boundary conditions
\begin{eqnarray}
V(T,y,{\bf x})=\big(\frac{y}{T}-1\big)^{+},
\end{eqnarray}
\begin{eqnarray}
V(t,0,{\bf x})=0.
\end{eqnarray}
For each $t\in {\cal T}$ and $i=1,2,...,N,$ let $V_{i}=V(t,y,e_{i})$ and ${\bf V}:=(V_{1},V_{2},...,V_{N}).$ We have that ${\bf V}$ satisfies the system of coupled PDEs
\begin{eqnarray}
\frac{\partial V_{i}}{\partial t}+(1-r_{i}y)\frac{\partial V_{i}}{\partial y}+\frac{1}{2}\sigma_{i}^{2}y^{2} \frac{\partial^{2}V_{i}}{\partial y^{2}}+ \left <{\bf V}, {\bf \tilde{A}} e_{i} \right > = 0 \ ,
\end{eqnarray}
and the boundary conditions
\begin{eqnarray}
V(T,y,e_{i})=\big(\frac{y}{T}-1\big)^{+},
\end{eqnarray}
\begin{eqnarray}
V(t,0,e_{i})=0,
\end{eqnarray}
for each $i=1,2,...,N.$
\section{A closed-form formula}
\renewcommand{\theequation}{4.\arabic{equation}}
\setcounter{equation}{0}
In this section, we restrict ourselves to a special case with the number of regimes $N$ being $2$ in order
to simplify our discussion. By means of the homotopy analysis method, we derive a closed-form solution for a floating strike Asian option under a regime switching model.
To solve the system of PDEs effectively, we shall introduce
the transformations $z=-\ln(y)$ and $\tau_{i}=(T-t)\frac{\sigma_{i}^{2}}{2}, i=1,2.$ Then the system of equations (3.15)-(3.17) becomes
\begin{eqnarray}
\left\{\begin{array}{lll}
{\cal L}_{1} V_{1}(\tau_{1},z)= \lambda_{1}\big(V_{1}(\tau_{1},z)-V_{2}(\tau_{2},z)\big)-\frac{2e^{z}}{\sigma_{1}^{2}}\frac{\partial V_{1}}{\partial z}\,
 \\
 V_{1}(0,z)=0 \,
 \\
\lim_{z\to \infty}V_{1}(\tau_{1},z)=0 \,
\end{array} \right.
\end{eqnarray}
where
\begin{eqnarray}
 {\cal L}_{1}=\frac{\partial}{\partial \tau_{1}}-\frac{\partial^{2}}{\partial z^{2}}-(1+\gamma_{1})\frac{\partial }{\partial z} \,
\end{eqnarray}
and
\begin{eqnarray}
\left\{\begin{array}{lll}
 {\cal L}_{2} V_{2}(\tau_{2},z)=\lambda_{2}\big(V_{2}(\tau_{2},z)-V_{1}(\tau_{1},z)\big)-\frac{2e^{z}}{\sigma_{2}^{2}}\frac{\partial V_{2}}{\partial z}\,
 \\
 V_{2}(0,z)=0 \,
 \\
\lim_{z\to \infty}V_{2}(\tau_{2},z)=0 \,
\end{array} \right.
\end{eqnarray}
where
\begin{eqnarray}
 {\cal L}_{2}=\frac{\partial}{\partial \tau_{2}}-\frac{\partial^{2}}{\partial z^{2}}-(1+\gamma_{2})\frac{\partial }{\partial z},
\end{eqnarray}
$\lambda_{i}=\frac{2a_{ii}}{\sigma_{i}^{2}}$ and $\gamma_{i}=\frac{2r_{i}}{\sigma_{i}^{2}}, i=1,2.$
 The homotopy analysis method is adopted to solve $ V_{i}, i=1,2$ from equations (4.1) and (4.3).
Now we introduce an embedding parameter $p\in [0, 1]$ and construct unknown functions
${\bar{V}}_{i}(\tau_{i},z,p), i=1, 2$ that satisfy the following differential systems:
\begin{eqnarray}
\left\{\begin{array}{lll}
(1-p) {\cal L}_{1}[{\bar{V}}_{1}(\tau_{1},z,p)-{\bar{V}}_{1}^{0}(\tau_{1},z)]=-p\bigg\{{\cal A}_{1}[{\bar{V}}_{1}(\tau_{1},z,p), {\bar{V}}_{2}(\tau_{2},z,p)]  \bigg\} \,
 \\
 {\bar{V}}_{1}(0,z, p)= (1-p){\bar{V}}_{1}^{0}(0,z) \,
\\
\lim_{z\to \infty}{\bar{V}}_{1}(\tau_{1}, z, p)=0 \,
\end{array} \right.
\end{eqnarray}
\begin{eqnarray}
\left\{\begin{array}{lll}
(1-p) {\cal L}_{2}[{\bar{V}}_{2}(\tau_{2},z,p)-{\bar{V}}_{2}^{0}(\tau_{2},z)]=-p\bigg\{{\cal A}_{2}[{\bar{V}}_{1}(\tau_{1},z,p), {\bar{V}}_{2}(\tau_{2},z,p)]  \bigg\} \,
 \\
 {\bar{V}}_{2}(0,z, p)= (1-p){\bar{V}}_{2}^{0}(0,z) \,
\\
\lim_{z\to \infty}{\bar{V}}_{2}(\tau_{2}, z, p)=0 \,
\end{array} \right.
\end{eqnarray}
Here ${\cal L}_{i}, i=1,2$ is a differential operator defined as
\begin{eqnarray}
{\cal L}_{i}=\frac{\partial}{\partial \tau_{i}}-\frac{\partial^{2}}{\partial z^{2}}-(1+\gamma_{i})\frac{\partial }{\partial z}  \,
\end{eqnarray}
and ${\cal A}_{i}, i=1,2$ are functionals defined as
\begin{eqnarray}
&&{\cal A}_{1}[{\bar{V}}_{1}(\tau_{1},z,p),{\bar{V}}_{2}(\tau_{2},z,p)]\nonumber\\
&=&{\cal L}_{1}[{\bar{V}}_{1}(\tau_{1},z,p)]-\lambda_{1}({\bar{V}}_{1}(\tau_{1},z,p)-{\bar{V}}_{2}(\tau_{2},z,p))+\frac{2e^{z}}{\sigma_{1}^{2}}\frac{\partial {\bar{V}}_{1}}{\partial z}(\tau_{1},z,p),
\end{eqnarray}
\begin{eqnarray}
&&{\cal A}_{2}[{\bar{V}}_{1}(\tau_{1},z,p),{\bar{V}}_{2}(\tau_{2},z,p)]\nonumber\\
&=& {\cal L}_{2}[{\bar{V}}_{2}(\tau_{2},z,p)]-\lambda_{2} ({\bar{V}}_{2}(\tau_{2},z,p)-{\bar{V}}_{1}(\tau_{1},z,p))+\frac{2e^{z}}{\sigma_{2}^{2}}\frac{\partial {\bar{V}}_{2}}{\partial z}(\tau_{2},z,p).
\end{eqnarray}
With $p=1$, we have
\begin{eqnarray}
\left\{\begin{array}{lll}
{\cal L}_{1}[{\bar{V}}_{1}(\tau_{1},z,1)]=\lambda_{1}({\bar{V}}_{1}(\tau_{1},z,1)-{\bar{V}}_{2}(\tau_{2},z,1))-\frac{2e^{z}}{\sigma_{1}^{2}}\frac{\partial {\bar{V}}_{1}}{\partial z}(\tau_{1},z,1) \,
\\
 {\bar{V}}_{1}(0,z, 1)=0 \,
\\
\lim_{z\to \infty}{\bar{V}}_{1}(\tau_{1}, z, 1)=0 \,
\end{array} \right.
\end{eqnarray}
\begin{eqnarray}
\left\{\begin{array}{lll}
{\cal L}_{2}[{\bar{V}}_{2}(\tau_{2},z,1)]=\lambda_{2} ({\bar{V}}_{2}(\tau_{2},z,1)-{\bar{V}}_{1}(\tau_{1},z,1))-\frac{2e^{z}}{\sigma_{2}^{2}}\frac{\partial {\bar{V}}_{2}}{\partial z}(\tau_{2},z,1) \,
 \\
 {\bar{V}}_{2}(0,z, 1)=0 \,
\\
\lim_{z\to \infty}{\bar{V}}_{2}(\tau_{2}, z, 1)=0 \,
\end{array} \right.
\end{eqnarray}
Comparing with equations (4.1) and (4.3), it is obvious that ${\bar{V}}_{i}(\tau_{i},z,1), i=1,2$ are equal to our searched solutions $V_{i}(\tau_{i},z), i=1,2$.

Now we set $p=0$, equations (4.5) and (4.6) become
\begin{eqnarray}
\left\{\begin{array}{lll}
 {\cal L}_{1}[{\bar{V}}_{1}(\tau_{1},z,0)]={\cal L}_{1}[{\bar{V}}_{1}^{0}(\tau_{1},z)] \,
 \\
 {\bar{V}}_{1}(0, z, 0)={\bar{V}}_{1}^{0}(0,z) \,
 \\
\lim_{z\to \infty}{\bar{V}}_{1}(\tau_{1}, z, 0)=0 \,
\end{array} \right.
\end{eqnarray}
\begin{eqnarray}
\left\{\begin{array}{lll}
 {\cal L}_{2}[{\bar{V}}_{2}(\tau_{2},z,0)]={\cal L}_{2}[{\bar{V}}_{2}^{0}(\tau_{2},z)] \,
 \\
 {\bar{V}}_{2}(0, z, 0)={\bar{V}}_{2}^{0}(0,z) \,
 \\
\lim_{z\to \infty}{\bar{V}}_{2}(\tau_{2}, z, 0)=0 \,
\end{array} \right.
\end{eqnarray}
Clearly ${\bar{V}}_{i}(\tau_{i},z,0), i=1,2$ will be equal to ${\bar{V}}_{i}^{0}(\tau_{i},z)$ as long as the initial guess satisfies the limiting condition $\lim_{z\to \infty}{\bar{V}}_{i}^{0}(\tau_{i}, z)=0$. The limiting condition is the only requirement for ${\bar{V}}_{i}^{0}(\tau_{i}, z)$. However if in addition ${\cal L}_{i}[{\bar{V}}_{i}^{0}(\tau_{i},z)]=0$ will speed up a convergence rate of the solution series.

For the choice of ${\bar{V}}_{i}^{0}(\tau_{i},z)$ any continuous function satisfying the limiting condition can be used. We choose the corresponding European option value as the initial guess with two apparent merits given as in Zhu (2006):
\begin{enumerate}
\item {the boundary condition as $z \to \infty$ is automatically satisfied;}\\
\item {${\cal L}_{i}[{\bar{V}}_{i}^{0}(\tau_{i}, z)]$ will become $0$; which we expect will foster a faster convergence of the series.}
\end{enumerate}
 The closed-form solution of a European put option with a two-state regime switching is given in Zhu, et al. (2012):
\begin{eqnarray}
&&{V}_{i}^{0}(t,S)\nonumber\\
&=&Ke^{-r_{i}(T-t)}+\frac{1}{4 \pi \sqrt{2}}\sqrt{SK}e^{-\frac{1}{2}(r_{i}+a_{21}+a_{12}+\frac{\sigma_{1}^{2}+\sigma_{2}^{2}}{8})(T-t)}\int_{0}^{\infty}\frac{(-1)^{i-1} 2 {\hat{f}}_{1}(\rho)(a_{21}+a_{12})}{M(\rho)(\rho^{4}+\frac{1}{16})(\sigma_{1}^{2}-\sigma_{2}^{2})}\nonumber\\
&\times& \bigg\{e^{X_{i}(\rho)}\bigg[(2\rho^{2}-\frac{1}{2})\sin({\hat{f}}_{2}(\rho)+\theta(\rho)-{\hat{Y}}_{i}(\rho))-(2\rho^{2}+\frac{1}{2})\cos( {\hat{f}}_{2}(\rho)+\theta(\rho)-{\hat{Y}}_{i}(\rho) )   \bigg]\nonumber\\
&-& e^{-X_{i}(\rho)}\bigg[(2\rho^{2}-\frac{1}{2})\sin({\hat{f}}_{2}(\rho)+\theta(\rho)+{\hat{Y}}_{i}(\rho))-(2\rho^{2}+\frac{1}{2})\cos( {\hat{f}}_{2}(\rho)+\theta(\rho)+{\hat{Y}}_{i}(\rho) ) \bigg] \bigg\}\nonumber\\
&+&\frac{2{\hat{f}}_{1}(\rho)}{M(\rho)}\bigg\{e^{X_{i}(\rho)}\big[\sin({\hat{f}}_{2}(\rho)+\theta(\rho)-{\hat{Y}}_{i}(\rho))+\cos( {\hat{f}}_{2}(\rho)+\theta(\rho)-{\hat{Y}}_{i}(\rho) )\big]\nonumber\\
&-&e^{-X_{i}(\rho)} \bigg[\sin({\hat{f}}_{2}(\rho)+\theta(\rho)+{\hat{Y}}_{i}(\rho))+\cos( {\hat{f}}_{2}(\rho)+\theta(\rho)+{\hat{Y}}_{i}(\rho) )\bigg] \bigg\}\nonumber\\
&+& \frac{{\hat{f}}_{1}(\rho)}{\rho^{4}+\frac{1}{16}}\bigg\{e^{X_{i}(\rho)}\bigg[ (2\rho^{2}-\frac{1}{2})\sin({\hat{f}}_{2}(\rho)-{\hat{Y}}_{i}(\rho))-(2\rho^{2}+\frac{1}{2})\cos( {\hat{f}}_{2}(\rho)-{\hat{Y}}_{i}(\rho) ) \bigg]\nonumber\\
&+&e^{-X_{i}(\rho)}\bigg[(2\rho^{2}-\frac{1}{2})\sin({\hat{f}}_{2}(\rho)+{\hat{Y}}_{i}(\rho))-(2\rho^{2}+\frac{1}{2})\cos( {\hat{f}}_{2}(\rho)+{\hat{Y}}_{i}(\rho) ) \bigg]    \bigg\} d\rho, \nonumber\\
\end{eqnarray}
for $i=1,2,$ where
$$\tau=\frac{\sigma_{1}^{2}-\sigma_{2}^{2}}{4}(T-t),\quad \alpha=\frac{2(a_{12}-a_{21})}{\sigma_{1}^{2}-\sigma_{2}^{2}}, \quad \mu^{2}=\frac{4 a_{12}a_{21}}{(\sigma_{1}^{2}-\sigma_{2}^{2})^{2}},      $$
$$M(\rho)=\bigg\{\big[(\frac{1}{4}+\alpha)^{2}-\rho^{4}+\mu^{2}\big]^{2}+4 \rho^{4}(\frac{1}{4}+\alpha)^{2} \bigg\}^{\frac{1}{4}}, $$
$$\theta(\rho)=\frac{1}{2} \tan^{-1}\bigg[\frac{2\rho^{2}(\frac{1}{4}+\alpha)}{(\frac{1}{4}+\alpha)^{2}-\rho^{4}+\mu^{2}} \bigg],   $$
$$X_{i}(\rho)=(-1)^{i-1}M(\rho)\tau \cos\theta(\rho), \quad {\hat{Y}}_{i}(\rho)=(-1)^{i-1}M(\rho)\tau \sin\theta(\rho)     $$
and
$${\hat{f}}_{1}(\rho)=e^{-\frac{\rho}{\sqrt{2}}|\ln(\frac{S}{K})+r_{i}(T-t)|}, \quad {\hat{f}}_{2}(\rho)=\frac{\rho^{2}}{4}(\sigma_{1}^{2}+\sigma_{2}^{2})(T-t)-\frac{\rho}{\sqrt{2}}|\ln(\frac{S}{K})+r_{i}(T-t)|.$$
To find the values of ${\bar{V}}_{i}(\tau_{i}, z, 1), i=1,2$,
 we can expand the functions ${\bar{V}}_{i}(\tau_{i}, z, p)$ as a Taylor's series expansion of $p$
 \begin{eqnarray}
\left\{\begin{array}{lll}
{\bar{V}}_{1}(\tau_{1}, z, p)=\sum_{m=0}^{\infty}\frac{{\bar{V}}_{1}^{m}(\tau_{1}, z)}{m !}p^{m}\,
\\
{\bar{V}}_{2}(\tau_{2}, z, p)=\sum_{m=0}^{\infty}\frac{{\bar{V}}_{2}^{m}(\tau_{2}, z)}{m !}p^{m}\,
\\
\end{array} \right.
\end{eqnarray}
 where
 \begin{eqnarray}
\left\{\begin{array}{lll}
{\bar{V}}_{1}^{m}(\tau_{1}, z)=\frac{\partial ^{m}}{\partial p^{m}} {\bar{V}}_{1}(\tau_{1}, z, p)|_{p=0}\,
\\
{\bar{V}}_{2}^{m}(\tau_{2}, z)=\frac{\partial ^{m}}{\partial p^{m}} {\bar{V}}_{2}(\tau_{2}, z, p)|_{p=0}\,
\\
\end{array} \right.
\end{eqnarray}
To find ${\bar{V}}_{1}^{m}(\tau_{1}, z)$ and ${\bar{V}}_{2}^{m}(\tau_{2}, z)$ in equation (4.15), we put (4.15) into equations (4.5) and (4.6) respectively and obtain the following recursive relations:
\begin{eqnarray}
\left\{\begin{array}{lll}
{\cal L}_{1}({\bar{V}}_{1}^{m})=\lambda_{1}({\bar{V}}_{1}^{m-1}-{\bar{V}}_{2}^{m-1})-\frac{2e^{z}}{\sigma_{1}^{2}}\frac{\partial {\bar{V}}_{1}^{m-1}}{\partial z}\quad m=1,2,..., \,
 \\
 {\bar{V}}_{1}^{m}(0, z)=0 \,
 \\
\lim_{z\to \infty}{\bar{V}}_{1}^{m}(\tau_{1}, z)=0 \,
\\
\end{array} \right.
\end{eqnarray}
\begin{eqnarray}
\left\{\begin{array}{lll}
{\cal L}_{2}({\bar{V}}_{2}^{m})=\lambda_{2} ({\bar{V}}_{2}^{m-1}-{\bar{V}}_{1}^{m-1})-\frac{2e^{z}}{\sigma_{2}^{2}}\frac{\partial {\bar{V}}_{2}^{m-1}}{\partial z}\quad m=1,2,..., \,
 \\
 {\bar{V}}_{2}^{m}(0, z)=0 \,
 \\
\lim_{z\to \infty}{\bar{V}}_{2}^{m}(\tau_{2}, z)=0 \,
\\
\end{array} \right.
\end{eqnarray}
We introduce the following transformation:
\begin{eqnarray}
{\bar{V}}_{i}^{m}(\tau_{i}, z)=e^{-[(1+\gamma_{i})\frac{z}{2}+\frac{1}{4}\tau_{i}(1+\gamma_{i})^{2}]}{\hat{V}}_{i}^{m}(\tau_{i}, z).
\end{eqnarray}
 We can rewrite equations (4.17) and (4.18) in the form of standard nonhomogeneous diffusion equations
 \begin{eqnarray}
\left\{\begin{array}{lll}
\frac{\partial {\hat{V}}_{1}^{m}}{\partial \tau_{1}}-\frac{\partial^{2} {\hat{V}}_{1}^{m}}{\partial z^{2}}=e^{[(1+\gamma_{1})\frac{z}{2}+\frac{1}{4}\tau_{1}(1+\gamma_{1})^{2}]} \big[\lambda_{1}({\bar{V}}_{1}^{m-1}-{\bar{V}}_{2}^{m-1})-\frac{2e^{z}}{\sigma_{1}^{2}}\frac{\partial {\bar{V}}_{1}^{m-1}}{\partial z}\big] \,
 \\
 {\hat{V}}_{1}^{m}(0, z)=0 \,
 \\
\lim_{z\to \infty}{\hat{V}}_{1}^{m}(\tau_{1}, z)=0 \,
\\
\end{array} \right.
\end{eqnarray}
\begin{eqnarray}
\left\{\begin{array}{lll}
\frac{\partial {\hat{V}}_{2}^{m}}{\partial \tau_{2}}-\frac{\partial^{2} {\hat{V}}_{2}^{m}}{\partial z^{2}}=e^{[(1+\gamma_{2})\frac{z}{2}+\frac{1}{4}\tau_{2}(1+\gamma_{2})^{2}]} \big[\lambda_{2}({\bar{V}}_{2}^{m-1}-{\bar{V}}_{1}^{m-1})-\frac{2e^{z}}{\sigma_{2}^{2}}\frac{\partial {\bar{V}}_{2}^{m-1}}{\partial z}\big]
 \\
 {\hat{V}}_{2}^{m}(0, z)=0 \,
 \\
\lim_{z\to \infty}{\hat{V}}_{2}^{m}(\tau_{2}, z)=0 \,
\\
\end{array} \right.
\end{eqnarray}
The system of PDEs (4.20) and (4.21) has a well-known closed-form solution respectively (see Carslaw and Jaeger (1959)):
\begin{eqnarray}
 {\hat{V}}_{1}^{m}(\tau_{1}, z)&=& \int_{0}^{\tau_{1}}\int_{0}^{\infty} e^{[(1+\gamma_{1})\frac{\xi}{2}+\frac{1}{4}\tau_{1}(1+\gamma_{1})^{2}]}\nonumber\\
 &\times& \big[\lambda_{1}({\bar{V}}_{1}^{m-1}-{\bar{V}}_{2}^{m-1})-\frac{2e^{\xi}}{\sigma_{1}^{2}}\frac{\partial {\bar{V}}_{1}^{m-1}}{\partial \xi}\big]G_{1}(\tau_{1}-u,z,\xi)d\xi du,\nonumber\\
\end{eqnarray}
\begin{eqnarray}
 {\hat{V}}_{2}^{m}(\tau_{2}, z)&=&\int_{0}^{\tau_{2}}\int_{0}^{\infty} e^{[(1+\gamma_{2})\frac{\xi}{2}+\frac{1}{4}\tau_{2}(1+\gamma_{2})^{2}]}\nonumber\\
 &\times& \big[\lambda_{2}({\bar{V}}_{2}^{m-1}-{\bar{V}}_{1}^{m-1})-\frac{2e^{\xi}}{\sigma_{2}^{2}}\frac{\partial {\bar{V}}_{2}^{m-1}}{\partial \xi}\big]G_{2}(\tau_{2}-u,z,\xi)d\xi du,\nonumber\\
\end{eqnarray}
where
\begin{eqnarray}
G_{1}(\tau_{1},z,\xi)&=&\frac{1}{2\sqrt{\pi \tau_{1}}}\bigg\{\exp\bigg[\frac{-(z-\xi)^{2}}{4\tau_{1}} \bigg]+\exp\bigg[\frac{-(z+\xi)^{2}}{4\tau_{1}} \bigg]\nonumber\\
&+&(1-\gamma_{1})\sqrt{\pi \tau_{1}}\exp\bigg[\frac{(1-\gamma_{1})^2}{4}-\frac{(1-\gamma_{1})(z+\xi)}{2}\bigg]\nonumber\\
&\times& \operatorname{erfc}\bigg(\frac{z+\xi}{2\sqrt{\tau_{1}}}-\frac{1}{2}(1-\gamma_{1})\sqrt{\tau_{1}}\bigg)\bigg\},
\end{eqnarray}
\begin{eqnarray}
G_{2}(\tau_{2}, z,\xi)&=&\frac{1}{2\sqrt{\pi \tau_{2}}}\bigg\{\exp\bigg[\frac{-(z-\xi)^{2}}{4\tau_{2}} \bigg]+\exp\bigg[\frac{-(z+\xi)^{2}}{4\tau_{2}} \bigg]\nonumber\\
&+&(1-\gamma_{2})\sqrt{\pi \tau_{2}}\exp\bigg[\frac{(1-\gamma_{2})^2}{4}-\frac{(1-\gamma_{2})(z+\xi)}{2}\bigg]\nonumber\\
&\times& \operatorname{erfc}\bigg(\frac{z+\xi}{2\sqrt{\tau_{2}}}-\frac{1}{2}(1-\gamma_{2})\sqrt{\tau_{2}}\bigg)\bigg\},
\end{eqnarray}
and $\operatorname{erfc}(.)$ denotes the complementary error function.
\section{A symmetry between fixed-strike and floating-strike Asian options}
\renewcommand{\theequation}{5.\arabic{equation}}
\setcounter{equation}{0}
 There are many known symmetry results in financial option pricing. Henderson and Wojakowski (2002) derived an equivalence of European floating strike Asian calls (or put) and fixed strike Asian puts (or call) in the Black-Scholes-Merton model. Eberlein and Papapantoleon (2005) established a symmetry relationship between floating strike and fixed strike Asian options for
assets driven by general L\'{e}vy processes using a change of num\'{e}raire and the characteristic triplet of the dual process.

We also have a symmetry between fixed-strike and floating-strike Asian options in the regime switching model.
 In the framework of the regime switching model, the symmetry results for Asian options with respect to
${\cal G}_t$ are stated in the following proposition.

\noindent {\bf Proposition 5.1:} Let $C_{f}$($P_{f}$) denote the floating-strike Asian call (put) option and $C_{x}$ ($P_{x}$) the fixed strike Asian call (put) option; based on arithmetic averaging, written at time $t=0$ with expiration date $T$. Then the following symmetry results hold:
\begin{eqnarray}
C_{f}(S_{0},1,r,0,0,T)=P_{x}(S_{0},S_{0},0,r,0,T),\\
C_{x}(K,S_{0},r,0,0,T)=P_{f}(S_{0},\frac{K}{S_{0}},0,r,0,T).
\end{eqnarray}
The proof of Proposition 5.1 is similar to the proof of Theorem 1
in Henderson and Wojakowski (2002). Hence, we do not repeat it here.
\section{Conclusion}
We consider the pricing of a floating-strike Asian option in a two-state regime switching model.
 A closed-form analytical pricing formula for the floating strike Asian option is derived by the means of the homotopy analysis method.


\begin{thebibliography}{9}

\bibitem{BS}  Black F. and Scholes M., The pricing of options and corporate liabilities, \textsl{Journal of Political Economy,} \textbf{81
}, 1973, pp. 637--659.

\bibitem{BD} Boyle P., Broadie M., Glasserman P., Monte Carlo methods for security
pricing, \textsl{Journal of Economic Dynamics and Control,} \textbf{21},1997, pp. 1267--1321.

\bibitem{BD} Boyle, P. P., and Draviam T., Pricing exotic options under regime switching, \textsl{Insurance: Mathematics and Economics,} \textbf{40}, 2007,
pp. 267--282.

\bibitem{BE} Buffington J. and Elliott R. J., Regime
switching and European options. In: Stochastic Theory and
Control, Proceedings of a Workshop, Lawrence, K.S., pp.73-81.
Berlin: Springer 2002a

\bibitem{BE} Buffington J. and Elliott R.J., American
options with regime switching, \textsl{ International Journal of
Theoretical and Applied Finance,} \textbf{5
}, 2002b, pp. 497-514.

\bibitem{CJ59} Carslaw H.S. and Jaeger J.C., Conduction of heat in solids,
Clarendon Press: Oxford 1959.

\bibitem{C} Curran V., Valuing Asian options and portfolio options by conditioning on
the geometric mean price,  \textsl{Management Science,} \textbf{40}, 1994, pp.1705--1711.

\bibitem{C} Eberlein E. and Papapantoleon A., Equivalence of floating and fixed strike Asian and lookback options, \textsl{ Stochastic Processes and their Applications,} \textbf{115},
2005, pp. 31--40.

\bibitem{EAM} Elliott R. J., Aggoun L. and Moore J. B.,
Hidden Markov models: estimation and control.
Berlin Heidelberg New York: Springer 1994

\bibitem{EH} Elliott R. J. and van der Hoek J., An
application of hidden Markov models to asset allocation
problems, \textsl{Finance and Stochastics,} \textbf{3}, 1997, pp. 229--238.

\bibitem{EHJ} Elliott R. J., Hunter W. C. and Jamieson B. M.,
Financial signal processing, \textsl{International Journal of
Theoretical and Applied Finance,} \textbf{4}, 2001, pp. 567--584.

\bibitem{EMT} Elliott R. J., Malcolm W. P. and Tsoi A. H.,
Robust parameter estimation for asset price models with Markov
modulated volatilities,\textsl{Journal of Economics Dynamics and
Control,} \textbf{27}(8), 2003, pp. 1391--1409.

\bibitem{ECS} Elliott, R. J., Chan L. L., and Siu T. K.,
Option pricing and Esscher transform under regime
switching, \textsl{ Annals of Finance,} \textbf{1}(4), 2005, pp. 423--432.

\bibitem{F} Fu M.C., Madan D.B., Wang T., Pricing continuous Asian options:
A comparison of Monte Carlo and Laplace transform inversion methods, \textsl{Journal of
Computational Finance,} \textbf{2}(2), 1999, pp. 49--74.

\bibitem{G} Geman H., Yor M., Bessel processes, Asian options and perpetuities,
\textsl{Mathematical Finance,} \textbf{3}, 1993, pp.349--375.

\bibitem{GO10} Gounden S. and O'Hara J.G., An analytic formula for the price of an American-style Asian option of floating strike type,
 \textsl{Applied Mathematics and Computation,} \textbf{ 217}, 2010, pp. 2923--2936.

\bibitem{G} Guo X., Information and option pricings,
 \textsl{Quantitative Finance,} \textbf{1}, 2001, pp. 38--44.

\bibitem{H} Henderson V.,  Wojakowski R., On the equivalence of floating-and fixed-strike Asian options,  \textsl{Journal of Applied Probability,} \textbf{39}, 2002, pp.391--394.

\bibitem{K} Kemna A.G.Z., Vorst E.C.F., A pricing method for options based on
average values, \textsl{Journal of Banking and Finance,} \textbf{14}, 1990, pp. 113--129.

\bibitem{L13} Leung  K. S., An analytic pricing formula for lookback options under stochastic volatility,
 \textsl{Applied Mathematics Letters,} \textbf{26}, 2013, pp. 145--149.

\bibitem{L97}  Liao S.-J., Numerically solving non-linear problems by
the homotopy analysis method, \textsl{Comput. Mech.,} \textbf{20}, 1997, pp. 530--540.

\bibitem{LC02}  Liao S.-J. and  Campo A.,  Analytic solutions of the
temperature distribution in Blasius viscous flow problems,
\textsl{J. Fluid Mech.,} \textbf{453}, 2002, pp. 411--425.

\bibitem{LZ96}  Liao S.-J. and Zhu J.-M., A short note on higher-order streamfunction-vorticity formulations of 2D steady state Navier-Stokes equations, \textsl{Int. J. Numer.
Methods Fluids,} \textbf{22}, 1996, pp. 1--9.

\bibitem{LZ99} Liao  S.-J. and Zhu S.-P., Solving the Liouville equation with the general boundary element method approach, \textsl{Boundary Element Technology XIII,} 1999,
pp. 407--416.

\bibitem{Merton}  Merton R., Theory of rational option pricing, \textsl{Bell Journal of Economics and Management Science}, \textbf{4}, 1973, pp. 141--183.

\bibitem{OR70}  Ortega J.M.,  Rheinboldt W.C., Iterative solution of nonlinear equations in several variables, Academic Press, New York, 1970.

\bibitem{PS06} Peskir G., Shiryaev A., Optimal stopping and free-boundary problems, Birkhäuser-Verlag, 2006.

\bibitem{R} Ritchken P., Sankarasubramanian L., Vijh A.M., The valuation of
path dependent contracts on the average, \textsl{Management Science,} \textbf{39}, 1993, pp. 1202--1213.

\bibitem{R} Rogers L.C.G., Shi Z., The value of an Asian option, \textsl{ J. Appl. Prob.,}
\textbf{32}, 1995, pp.1077--1088.

\bibitem{Siu} Siu, T. K., Fair valuation of participating policies with surrender options and regime switching, \textsl{Insurance: Mathematics and
Economics,}  \textbf{37} (3), 2005, pp. 533--552.

\bibitem{T} Turnbull S., Wakeman L., A quick algorithm for pricing European
average options, \textsl{ Journal of Financial and Quantitative Analysis,} \textbf{26}, 1991, pp. 377--389.

\bibitem{Y}  Yao D.D., Zhang Q. and Zhou X.Y., A regime switching model for European options,
Working paper, 2003, Columbia University, New York, USA.

\bibitem{Y} Yuen F. L. and Yang H., Option pricing in a jump-diffusion model with regime-switching, \textsl{ASTIN Bulletin,} \textbf{39} (2), 2009, pp. 515--539.

\bibitem{Y} Yuen F. L. and Yang H., Pricing Asian options and equity-indexed annuities with regime switching by the trinomial tree method,
\textsl{North American Actuarial Journal,} \textbf{14} (2), 2010, pp. 256--277.

\bibitem{Y} Zhang P. G., Flexible arithmetic Asian options, \textsl{The Journal of Derivatives,}
\textbf{2}, 1995, pp. 53--63.

\bibitem{zhu06}  Zhu S.-P., An exact and explicit solution for the valuation of American put options,  \textsl{Quantitative Finance,} \textbf{6}(3), 2006, pp. 229--242.

\bibitem{zhu12}  Zhu S.-P., Badran  A. and  Lu X., A new exact solution for pricing European options in a two-state
regime-switching economy, \textsl{ Computers and Mathematics with Applications,} \textbf{64}, 2012, pp. 2744--2755.
\end{thebibliography}
\end{document}